\documentstyle[emulateapj,apjfonts]{article}

\slugcomment{Accepted for publication in The Astrophysical Journal
Letters}

\begin{document}

\title{Recently discovered pulsars and unidentified EGRET sources}

\author{Diego F. Torres\footnote{Physics Department, Princeton University, NJ 08544},
 Yousaf M. Butt\footnote{Harvard-Smithsonian Center for Astrophysics, 60 Garden Street,
Cambridge, MA 02138}, and Fernando Camilo\footnote{Columbia
Astrophysics Laboratory, Columbia University, 550 West 120th
Street, New York, NY 10027}}

\begin{abstract}

We present a correlative study between all unidentified EGRET
sources at low Galactic latitudes and the newly discovered pulsars
in the released portion of the Parkes multibeam radio survey. We
note 14 positional coincidences: eight of these are ``Vela-like''
pulsars, with relatively small periods, small characteristic ages,
and high spin-down luminosities. Three of these coincidences have
been investigated by D'Amico et al. (2001) and Camilo et al.
(2001). Among the others, we argue that PSR~J1015$-$5719 may
plausibly generate part of the high energy radiation observed from
3EG~J1014$-$5705.  Three additional interesting cases are:
3EG~J1410$-$6147 and either of PSRs~J1412$-$6145 or J1413$-$6141,
if the pulsars are at the estimated distance of the coincident
SNR~G312.4$-$0.4; and 3EG~J1639$-$4702/PSR~J1637$-$4642.  The
remaining positional coincidences between the EGRET sources and
the newly discovered pulsars are almost certainly spurious.

\end{abstract}


\section{Introduction}

The Third EGRET Catalog lists 271 point sources (Hartman et al.
1999) of which about two thirds are not yet identified.  Although
most known young pulsars are not detected in the high-energy
window ($>100$\,MeV), there are so far six such confirmed
$\gamma$-ray pulsars (Thompson 1999).  They are relatively young
objects, with hard high-energy spectral indices, and apparently
showing a trend for spectral hardening with increasing
characteristic age (Fierro et al. 1993).  Young pulsars were
proposed as potential counterparts for $\gamma$-ray sources as
early as 1983 (D'Amico 1983; Helfand 1994), and their relationship
with unidentified EGRET sources remains a matter of current debate
(see, e.g., Zhang, Zhang, \& Cheng 2000).

In order to unambiguously identify a pulsar as the origin of the
$\gamma$-rays from an EGRET source, $\gamma$-ray pulsations must
be detected at the pulsar period.  However the $\gamma$-fluxes are
generally very low, and nearly contemporaneous radio/X-ray and
$\gamma$-ray observations are required to fold the few available
photons with the correct ephemeris.  Alas this is no longer
possible owing to the demise of the {\em Compton Gamma-Ray
Observatory\/}, and new candidate associations must for now be
judged by comparison with the properties of the six known EGRET
pulsars.

Supernova remnants (SNRs) interacting with nearby molecular clouds
have also been proposed as counterparts to some $\gamma$-ray
sources (e.g. Sturner, Dermer, \& Mattox 1996; Esposito et al.
1996; Combi et al. 1998a, b, 2001), and in such cases both a
pulsar and an associated interacting SNR may partially contribute
to generate a given EGRET source flux.

The Parkes multibeam pulsar survey is a large-scale survey of a
narrow strip of the inner Galactic plane ($|b|< 5\arcdeg$,
$260\arcdeg < l < 50\arcdeg$; Manchester et al. 2001). It has much
greater sensitivity than any previous survey to young and distant
pulsars along the Galactic plane, and it has resulted in the
detection of many previously unknown young pulsars, potentially
counterparts of EGRET sources. In this Letter we correlate the
released portion of the Parkes pulsar survey\footnote{See
http://www.atnf.csiro.au/research/pulsar/pmsurv.}, now consisting
of 368 pulsars, with all unidentified EGRET sources at low
latitudes, in order to determine whether there are any new likely
physical associations.

\section{Correlations}

Table~1 lists some important properties for the known EGRET
$\gamma$-ray pulsars.  Two of the parameters listed are highly
uncertain: the distance $d$, and the {\em observed\/} efficiency
for conversion of spin-down luminosity $\dot E$ into $\gamma$-ray
luminosity $L_{\gamma}$: $\eta \equiv L_{\gamma}/\dot E = f 4 \pi
d^2 F_{\gamma}/\dot E$, where $F_{\gamma}$ is the observed
$\gamma$-ray flux and $f$ is the $\gamma$-ray beaming fraction ($0
< f \le 1$). Because this fraction is essentially unknown (see,
e.g., Yadigaroglu \& Romani 1995), and in part because $f \sim 1$
results in $\eta \sim 100\%$ for at least one pulsar at its
nominal distance, it is common practice, which we follow, to {\em
assume\/} $f \equiv 1/4\pi$.  The efficiencies observed span the
range $0.01\% \la \eta \la 3$/19\%: the two upper limits for
$\eta$ stem from the different measurements of the distance to
PSR~B1055$-$52 (see \"{O}gelman \& Finley~1993; Combi et al.~1997;
Romero~1998).  We shall then consider $0.01\% \la \eta \la 10\%$
as ``reasonable'' values for evaluating the plausibility of new
associations, by analogy with presently known $\gamma$-ray
pulsars, even though this inferred range relies on an assumed
uniform value for $f$, and uncertain distances.

All positional coincidences between unidentified 3EG sources and
Parkes pulsars are given in Table~2.  We list for each 3EG source
the variability index $I$: pulsars are steady $\gamma$-ray
sources.  For each pulsar we also present the ``spin-down flux''
$\dot E/d^2$, which is generally well-correlated with the
detectability of $\gamma$-ray pulsars.  A strict (linear) reading
of this correlation would suggest $L_{\gamma} \propto \dot E$,
while Thompson et al. (1999) find that $L_{\gamma} \propto {\dot
E}^{1/2}$.  However, we summarily dismiss some potential
associations below owing to a low value of $\dot E/d^2$ only in
extreme cases where the difference between $\dot E$- and ${\dot
E}^{1/2}$-scaling is irrelevant (and where even a hefty error in a
more relevant and uncertain parameter --- distance --- does not
change matters significantly).  In all these discarded cases, in
addition, the efficiencies required would be $\eta \gg 1000\%$,
making the potential associations utterly unphysical.  Finally, in
evaluating possible associations, we compare the photon indices of
known pulsars (Table~1) with those of the EGRET sources (see,
e.g., Merck et al.  1996; Zhang \& Cheng 1998; Cheng \& Zhang
1998).

We have also quantified the chance probability for obtaining these
associations, adapting the numerical code described by Romero et
al. (1999a, b):  there is a 2\% probability of having 8 chance
coincidences between different unidentified 3EG sources and Parkes
pulsars, as in Table~2.\footnote{To arrive at this result we have
considered the 41 3EG sources in the range of the Parkes survey,
and simulated them in synthetic populations in order to obtain an
average expected number of chance coincidences.  There is
therefore a 98\% a priori probability that not all 8 associations
are spurious.}

Camilo et al.~(2001) proposed the possible physical association
\#1, with required efficiency of 0.5\% (Table~2).  This pulsar
appears to be located just outside the SNR~G284.3$-$1.8, which
itself is interacting with an adjacent molecular cloud (Ruiz \&
May 1986). Another Parkes pulsar, PSR~J1013$-$5934 (\#2 in
Table~2), is coincident with 3EG~1013$-$5915, but it is an old
pulsar ($\tau=12$\,Myr) with low $\dot E = 2.5\times
10^{32}$\,ergs\,s$^{-1}$ and cannot be a significant $\gamma$-ray
contributor. D'Amico et al.~(2001) have studied cases \#6 and 14
(efficiencies: $\eta = 2$\% and 7\%, respectively). Both are
plausible candidates to generate the respective EGRET source
fluxes. PSR~J1837$-$0559, is also coincident with 3EG~J1837$-$0606
(\#13), but its $\dot E/d^2$ is 80 times smaller than for
PSR~J1837$-$0604, and it cannot contribute significantly to the
$\gamma$-ray source. The pulsars in pairs \#8 and 9 have far too
low a spin-down luminosity at too large a distance (Table~2) to
explain their coincident $\gamma$-ray sources.  They are also both
old, with $\tau \sim 4$\,Myr.  Cases \#10 and 11 are discussed
elsewhere in connection with a proposal for a SNR shock origin of
the bulk of the $\gamma$-rays resulting from 3EG~J1714$-$3857
(Butt et al. 2001); neither of these pulsars is energetic enough
to contribute significant amounts of high-energy flux.

\section{Discussion}

We now discuss the remaining four EGRET sources positionally
superposed with five newly discovered pulsars. We provide
observational data on the apparent associations in Tables~3 and 4.
Cases \#3--5, 7 and 12 all contain Vela-like pulsars, with
relatively short periods, low characteristic ages, and high
spin-down luminosities $(\dot E \ga 10^{35}$\,ergs\,s$^{-1}$).

The pulsar in case \#3 would require an efficiency $\eta = 5\%$ at
its nominal distance to explain the luminosity of the
corresponding 3EG source, which has photon index 2.23 (see
Table~4). This spectrum is softer than that of the Crab, although
it is consistent with it within the uncertainties. It is also
consistent with the index of 3EG~2227+6122, for which
PSR~J2229+6114 has been proposed as the likely source (Halpern et
al. 2001).  The $\gamma$-ray source in case \#3 is not variable
(Tompkins 1999; Torres et al. 2001c), as expected from direct
pulsar or pulsar wind nebula/SNR shock emission. PSR~J1015$-$5719
has $\tau=39$\,kyr and $\dot E = 8.2\times
10^{35}$\,ergs\,s$^{-1}$ (Table~3).  While no cataloged SNR is
superposed with the 3EG source (Torres et al. 2001b), this absence
does not mean that one does not exist, and further sensitive
searches may prove fruitful.  Thus the connection between
3EG~J1014$-$5705 and PSR~J1015$-$5719 appears plausible and is
worth additional study.

The pulsars in pairs \#4 and 5 require unreasonably high
efficiencies at their nominal distances to explain the
$\gamma$-ray flux from the corresponding EGRET source ($\eta \ga
100$\%, Table~4).  However, both pulsars are located in the
direction of the Centaurus arm, and it is known that in such
directions the electron density/distance model of Taylor \& Cordes
(1993) can be unreliable, sometimes overestimating the distances
by factors of up to $\sim 4$ (see discussion in Camilo et al.
2001). Both pulsars are located (at least in projection) well
within the boundaries of the incomplete shell SNR~G312.4$-$0.4
(Caswell \& Barnes 1985), to which Yadigaroglu \& Romani (1997)
estimate a $\Sigma-D$ distance of 1.9\,kpc.  At this distance the
required efficiencies for the pulsars in cases \#4 and 5 would be
12\% and 3\%, respectively, which would make them considerably
more plausible sources of the observed high energy emission.
Furthermore, 3EG~J1410$-$6147 has a photon index comparable to
that of the Crab, and is not variable (Tables~1, 2 and 4).  Pairs
\#4 and 5 therefore appear intriguing.  However it should be noted
that $\Sigma-D$ distances are notoriously unreliable (e.g., for
this very SNR, Caswell \& Barnes 1985 and Case \& Bhattacharya
1999 infer values in substantial disagreement both with each other
and with that determined by Yadigaroglu \& Romani 1997).  Ideally,
further observations of SNR~G312.4$-$0.4 may indicate whether it
shows signs of interaction with PSRs~J1412$-$6145 or J1413$-$6141,
and possibly constrain their distances. Depending on the actual
distances, the $\gamma$-ray emission from 3EG~J1410$-$6147 may
conceivably arise from a combination of PSRs~J1412$-$6145,
J1413$-$6141, and/or SNR~G312.4$-$0.4.

The efficiency required to explain the EGRET flux in case \#7 is
$\eta = 12\%$, which seems possible.  However, the spectral index
of 2.50 is larger than those of known $\gamma$-ray pulsars.  One
of the SNRs coincident with the EGRET source, G337.8$-$0.1,
harbors a maser (Koralesky et al. 1998), which is indicative of
interaction between the SNR shock and the ambient medium.  Thus,
were a sufficiently massive molecular cloud located nearby, it
could help produce the high energy radiation as a result of
hadronic interaction (Aharonian, Drury, \& V\"olk 1994; Aharonian
\& Atoyan 1996).  Part of the EGRET flux could plausibly come from
PSR~J1637$-$4642 and part from pion $\gamma$-decay via
SNR~G337.8$-$0.1's interaction with the putative cloud.  In this
case, the photon index would reflect a weighted average value.
Both possible mechanisms for the high energy emission would
produce a non-variable source, as is the case for
3EG~J1639$-$4702.

Lastly we consider case \#12, for which the required efficiency is
high at the nominal pulsar distance and upper limit flux value,
$\eta = 55$\%. The pulsar is located, in projection, just outside
the plerionic SNR~G27.8+0.6, for which the distance is $\sim
2$\,kpc (Reich et al. 1984). Although the estimated ages are
comparable ($\tau = 52$\,kyr for the pulsar and $\sim 45$\,kyr for
the SNR), it seems unlikely that both objects are physically
associated, given the offset between the centrally peaked SNR
component and the pulsar (see Reich et al. 1984). Whatever the
possible relation between pulsar and SNR, the 3EG source in case
\#12 is variable (Table~2), arguing against a pulsar origin.

Examination of X-ray archives via HEASARC has revealed no
compelling counterpart sources to any of the pulsars in Table 3.
(A possible {\em Einstein\/} X-ray source at the edge of an IPC
field near the location of PSR~J1015$-$5719 [\#3] is likely an
artifact resulting from the large effective area correction used.)
However, the exposures may not have been deep enough to reveal
such X-ray emission, usually expected at the level $L_x \sim
10^{-3} \dot E$ (e.g., Becker \& Trumper 1997). Future
observations of these pulsar fields with {\em XMM-Newton\/} and
{\em Chandra\/} could thus be instructive.

In conclusion, we find that two recently discovered Parkes pulsars
(PSRs~J1015$-$5719 and J1637$-$4642) could plausibly generate at
least part of the $\gamma$-ray flux observed from two unidentified
EGRET sources (3EG~J1014$-$5705 and 3EG~J1639$-$4702,
respectively), and either of PSRs~J1412$-$6145 or J1413$-$6141
could be associated with 3EG~J1410$-$6147 if they are located
closer than their DM distances by a factor of $\sim 4$. Cases
\#3--5 and 7 represent promising targets for the forthcoming {\em
AGILE\/} and {\em GLAST\/} missions.

\acknowledgements

D.F.T. was supported by CONICET and Fundaci\'on Antorchas, and is
on leave from IAR, Argentina. Y.M.B. acknowledges the support of
the High Energy Astrophysics division at the CfA and the {\em
Chandra\/} project.  Partial support from the US DOE grant
DE-FG02-91ER40671 (D.F.T.) and NASA grants NAS8-39073 (Y.M.B.) and
NAG5-9095 (F.C.) is acknowledged. We thank D. Nice and G.~E.
Romero for comments.


\begin{deluxetable}{lrrlllll}
\tablecaption{$\gamma$-ray pulsars
detected by EGRET } \tablewidth{0pt} \tablehead{
\colhead{Pulsar/3EG Jsource} & \colhead{$P$} & \colhead{$\tau$} &
\colhead{$\dot E$} & \colhead{$d$} & \colhead{$F^{3EG}_\gamma
[\times 10^{-8}]$} & \colhead{$\gamma^{3EG}$} & \colhead{$\eta $}
\nl \colhead{} & \colhead{(ms)}                              &
\colhead{(kyr)} & \colhead{(ergs\,s$^{-1}$)}                  &
\colhead{(kpc)} & \colhead{(ph cm$^{-2}$ s$^{-1}$)}           &
\colhead{} & \colhead{$^{(100{\rm MeV}-}_{-10{\rm GeV})}$} \nl }

\startdata Crab/0534$-$2200     &  33 &   1.2 & $5.0\times
10^{38}$ & 2.0  &
  226.2$\pm$4.7  & 2.19$\pm$0.02 & 0.01\% \\
Vela/0834$-$4511     &  89 &  12.5 & $6.3\times 10^{36}$ & 0.25  &
  834.3$\pm$11.2 & 1.69$\pm$0.01 & 0.08\%  \\
B1951+32/\nodata     &  39 & 100.0 & $3.7\times 10^{36}$ & 2.4  &
  \nodata        & \nodata       & 0.3\%  \\
B1706$-$44/1710$-$4439 & 102 &  15.8 & $3.1\times 10^{36}$ & 1.8 &
  111.2$\pm$6.2  & 1.86$\pm$0.04 & 1\%  \\
Geminga/0633+1751    & 237 & 316.2 & $3.1\times 10^{34}$ & 0.16 &
  352.9$\pm$5.7  & 1.66$\pm$0.01 & 3\%  \\
B1055$-$52/1058$-$5234 & 197 & 501.1 & $3.1\times 10^{34}$ &
0.5/1.5 &
  33.3$\pm$3.8   & 1.94$\pm$0.10 & 2/19\%  \\

\enddata
\tablecomments{Pulsar parameters and distances (and $\eta$ in the
case of PSR~B1951+32 --- this pulsar is not a 3EG source) are
taken from Kaspi et al. (2000), excepting PSR~B1055$-$52, for
which we also consider a smaller value of distance (\"{O}gelman \&
Finley 1993; Combi et al. 1997), and Vela (Caraveo et al. 2001 and
references therein). $\tau= P/2\dot P$, and $\dot E=4\pi^2 I \dot
P /P^3$, with $I=10^{45}$\,g\,cm$^2$. The ``P1234'' $\gamma$-ray
fluxes and spectral indices are from the 3EG catalog (Hartman et
al. 1999), from which we have computed the $\eta$ values.}
\end{deluxetable}

\begin{deluxetable}{llllllll}
\tablecaption{Positional coincidences
between unidentified 3EG sources and Parkes pulsars}
 \tablewidth{0pt} \tablehead{
\colhead{3EG~J}  & \colhead{Note} &\colhead{$I$} &\colhead{PSR~J}
& \colhead{\#} & \colhead{SNRs} & \colhead{Ref.} &\colhead{$\dot
E/d^2$}  } \startdata

1013$-$5915 & C, em &1.6 & 1016$-$5857 & 1 & G284.3$-$1.8 & Camilo et al. (2001)& $2.9\times 10^{35}$ \\
            &       &&  1013$-$5934 & 2  & & & $2.0\times 10^{30}$ \\
1014$-$5705 & C, em &1.4& 1015$-$5719 & 3  & & & $3.4\times 10^{34}$ \\
1410$-$6147 & C     &1.2& 1412$-$6145 & 4  & G312.4$-$0.4 & & $1.4\times 10^{33}$ \\
            &       && 1413$-$6141 & 5  & & & $1.6\times 10^{33}$ \\
1420$-$6038 & C     &2.1& 1420$-$6048 & 6  & & D'Amico et al. (2001)& $1.7\times10^{35}$\\
1639$-$4702 & C, em & 1.9&1637$-$4642 & 7  & G337.8$-$0.1; G338.1+0.4; G338.3+0.0& & $1.9\times 10^{34}$ \\
            &       & &1640$-$4648 & 8  & & & $1.5\times 10^{32}$ \\
            &       & &1637$-$4721 & 9  & & & $3.1\times 10^{30}$ \\
1714$-$3857 & C, em &2.1& 1713$-$3844 & 10 & G348.5+0.0; G348.5+0.1; G347.3$-$0.5& Butt et al. (2001) & $4.0\times 10^{31}$ \\
            &       &&1715$-$3903 & 11 & & Butt et al. (2001) & $3.0\times10^{33}$ \\
1837$-$0423 & C     & 5.4&1838$-$0453 & 12 & G27.8+0.6 & & $1.2\times 10^{33}$ \\
1837$-$0606 & C, em & 2.4& 1837$-$0559 & 13 & & & $6.5\times 10^{32}$ \\
            &       & &1837$-$0604 & 14 & & D'Amico et al. (2001) & $5.2\times 10^{34}$ \\
\enddata
\tablecomments{A number (\#) is assigned to each pair for ease of
reference.  SNRs contained in Green's (2000) catalog coinciding
with the {\em EGRET\/} sources are noted (see Torres et al.
2001b).  Only in cases \#4, 5 and 12 do the pulsars also coincide
with the SNRs listed. ``C'' and ``em'' refer to the $\gamma$-ray
sources:  source confusion exists and sources are possibly
extended or multiple, respectively (Hartman et al. 1999). $I$ is
the variability index as in Torres et al. (2001a, c), where $I>5$
($<2$) represents a source whose flux presents variability levels
at least 8\,$\sigma$ (less than $2\,\sigma$) above those displayed
by confirmed pulsars. Pulsar ``spin-down flux'' $\dot E/d^2$ is in
units of ergs\,s$^{-1}$ kpc$^{-2}$.}
\end{deluxetable}

\begin{deluxetable}{lllrrrrrl}
\tablecaption{Observational parameters
for the pulsars in the candidate associations} \tablewidth{0pt}
\tablehead{ \colhead{Case} & \colhead{$\Delta\theta$ [deg]} &
\colhead{$\theta$ [deg]} & \colhead{ $(l,b)$} & \colhead{$d$
[kpc]} & \colhead{$\tau$ [kyr]}& \colhead{$P$ [ms]}&
\colhead{$\dot P$ [10$^{-15}$] }& \colhead{$\dot E$
[ergs\,s$^{-1}$] } } \startdata
\#3  & 0.30 &0.67 &283.09,$-$0.58 & 4.9 & 38.7 & 140 &  57.4 & $8.2\times 10^{35}$\\
\#4  & 0.15 & 0.36&312.32,$-$0.37 & 9.3 & 50.6 & 315 &  98.7 & $1.2\times 10^{35}$\\
\#5  & 0.28 &0.36 &312.46,$-$0.34 &11.0 & 13.5 & 286 & 333.4 & $5.7\times 10^{35}$\\
\#7  & 0.46 &0.56 &337.79,$+$0.31 & 5.8 & 41.2 & 154 &  59.2 & $6.4\times 10^{35}$\\
\#12 & 0.50 & 0.52& 27.07,$+$0.71 & 8.2 & 52.2 & 381 & 115.7 & $8.3\times 10^{34}$\\
\enddata
\tablecomments{$\Delta\theta$ is the angular distance between the
center of the 3EG source and the position of the respective
coincident pulsar.  $\theta$ is the effective 95\% confidence
level radius of the 3EG source error box (Hartman et al. 1999).
Pulsar parameters are taken from the Parkes survey database (see
footnote 1), and their distances $(d)$ are estimated from the
observed dispersion measure (Taylor \& Cordes 1993).}
\end{deluxetable}

\begin{deluxetable}{llllllr}
\tablecaption{Observed and computed
$\gamma$-ray emission
parameters for new candidate associations}
 \tablewidth{0pt}
\tablehead{ \colhead{Case}& \colhead{$F^{3EG}$ [$\times
10^{-8}$\,ph cm$^{-2}$ s$^{-1}$]} & \colhead{$F^{3EG}$
[ergs\,cm$^{-2}$ s$^{-1}$]} & \colhead{$\gamma^{3EG}$} &
\colhead{$L_{isotropic}$ [ergs s$^{-1}$]} & \colhead{$L_{beamed}$
[ergs\,s$^{-1}$]} & \colhead{$\eta $ } } \startdata
\#3 & 34.0$\pm$6.5 & 1.90 $\times 10^{-10}$ & 2.23$\pm$0.20 & 5.2 $\times 10^{35}$ & 4.1 $\times 10^{34}$ &   5\% \\

\#4 & 64.2$\pm$8.8 & 4.09 $\times 10^{-10}$ & 2.12$\pm$0.14 & 4.1 $\times 10^{36}$ & 3.2 $\times 10^{35}$ & 270/12\%\tablenotemark{a} \\

\#5 & 64.2$\pm$8.8 & 4.09 $\times 10^{-10}$ & 2.12$\pm$0.14 & 5.7 $\times 10^{36}$ & 4.5 $\times 10^{35}$ &  80/3\%\tablenotemark{a} \\

\#7 & 53.2$\pm$8.7 & 2.30 $\times 10^{-10}$ & 2.50$\pm$0.18 & 9.3 $\times 10^{35} $& 7.4 $\times 10^{34}$ &  12\% \\

\#12 & $<$19.1 & $<0.70 \times 10^{-10}$ & 2.71$\pm$0.44 & $<5.8 \times 10^{35} $& $<4.6 \times 10^{34}$ &  $<$55\% \\

\enddata
\tablecomments{The EGRET fluxes correspond to the ``P1234'' values
(Hartman et al. 1999). Luminosities, and efficiency values for all
pulsars, are given for the distance estimated from the dispersion
measure (Table 3).} \tablenotetext{a}{Second values of efficiency
correspond to a distance of 1.9\,kpc, one of those estimated for
the positionally coincident SNR~G312.4$-$0.4 (see text). }
\end{deluxetable}

\end{document}